\documentclass[12pt,preprint]{aastex}
\def\degpoint{\ifmmode ^{\rm{o}}\!. \else $^{\rm{o}}\!.$\fi}
\newcommand{\degrees}{$^{\rm{o}}$}
\newcommand{\ms}{\mbox{m\,s$^{-1}$}}

\newcommand{\rchi}{\chi^2_{\nu}}
\newcommand{\Msun}{\mbox{M$_{\odot}$}}

\newcommand{\Mjup}{\mbox{M$_{\rm Jup}$}}

\newcommand{\ltsimeq}{\raisebox{-0.6ex}{$\,\stackrel
         {\raisebox{-.2ex}{$\textstyle <$}}{\sim}\,$}}

\begin{document}

\title{A Detailed Analysis of the HD 73526 2:1 Resonant Planetary System }

\author{Robert A.~Wittenmyer\altaffilmark{1,2}, Xianyu 
Tan\altaffilmark{3,4}, Man Hoi Lee\altaffilmark{3,5}, Jonathan 
Horner\altaffilmark{1,2}, C.G.~Tinney\altaffilmark{1,2}, 
R.P.~Butler\altaffilmark{6}, G.S.~Salter\altaffilmark{1,2}, 
B.D.~Carter\altaffilmark{7}, H.R.A.~Jones\altaffilmark{8}, 
S.J.~O'Toole\altaffilmark{9}, J.~Bailey\altaffilmark{1,2}, 
D.~Wright\altaffilmark{1,2}, J.D. Crane\altaffilmark{10}, S.A. 
Schectman\altaffilmark{10}, P.~Arriagada\altaffilmark{6}, 
I.~Thompson\altaffilmark{10}, D.~Minniti\altaffilmark{11,12}, \& 
M.~Diaz\altaffilmark{11} }

\altaffiltext{1}{School of Physics, University of New South Wales, 
Sydney 2052, Australia}
\altaffiltext{2}{Australian Centre for Astrobiology, University of New 
South Wales, Sydney 2052, Australia}
\altaffiltext{3}{Department of Earth Sciences, The University of Hong Kong,
Pokfulam Road, Hong Kong }
\altaffiltext{4}{Department of Planetary Sciences and Lunar and 
Planetary Laboratory, The University of Arizona, 1629 University 
Boulevard, Tucson, AZ 85721, USA }
\altaffiltext{5}{Department of Physics, The University of Hong Kong, 
Pokfulam Road, Hong Kong }
\altaffiltext{6}{Department of Terrestrial Magnetism, Carnegie
Institution of Washington, 5241 Broad Branch Road, NW, Washington, DC
20015-1305, USA}
\altaffiltext{7}{Faculty of Sciences, University of Southern Queensland,
Toowoomba, Queensland 4350, Australia}
\altaffiltext{8}{University of Hertfordshire, Centre for Astrophysics
Research, Science and Technology Research Institute, College Lane, AL10
9AB, Hatfield, UK}
\altaffiltext{9}{Australian Astronomical Observatory, PO Box 915,
North Ryde, NSW 1670, Australia}
\altaffiltext{10}{The Observatories of the Carnegie Institution of 
Washington, 813 Santa Barbara Street, Pasadena, CA 91101, USA}
\altaffiltext{11}{Institute of Astrophysics, Pontificia Universidad 
Catolica de Chile, Casilla 306, Santiago 22, Chile}
\altaffiltext{12}{Vatican Observatory, V00120 Vatican City State, Italy }

\email{
rob@phys.unsw.edu.au}

\shortauthors{Wittenmyer et al.}

\begin{abstract}

\noindent We present six years of new radial-velocity data from the 
Anglo-Australian and Magellan Telescopes on the HD\,73526 2:1 resonant 
planetary system.  We investigate both Keplerian and dynamical 
(interacting) fits to these data, yielding four possible configurations 
for the system.  The new data now show that both resonance 
angles are librating, with amplitudes of 40\degrees\ and 60\degrees, 
respectively.  We then perform long-term dynamical stability tests to 
differentiate these solutions, which only differ significantly in the 
masses of the planets.  We show that while there is no clearly preferred 
system inclination, the dynamical fit with $i=90$\degrees\ provides the 
best combination of goodness-of-fit and long-term dynamical stability.

\end{abstract} 

\keywords{planetary systems: individual (HD 73526) -- techniques: radial 
velocities -- methods: N-body simulations }

\section{Introduction}

The ever-growing population of known multiple-planet systems has proven 
to be an exceedingly useful laboratory for testing models of planetary 
system formation and dynamical evolution.  Of particular interest are 
the systems which are in, or near, resonant configurations.  A number of 
such systems have been identified from radial-velocity surveys, with 
some notable examples including GJ\,876 \citep{marcy01}, HD\,128311 
\citep{vogt05}, HD\,82943 \citep{mayor04}, and HD\,200964 
\citep{johnson11}.  \citet{wright11b} noted that about 1/3 of 
well-characterized multi-planet systems were in such low-order period 
commensurabilities.  The \textit{Kepler} mission has revealed hundreds 
of candidate multiply-transiting planetary systems \citep{borucki10, 
batalha13}, some of which are in or near low-order resonances 
\citep{lissauer11, steffen13}.  One emerging trend from the 
\textit{Kepler} results is that a significant number of such 
near-resonant planet pairs are outside of the resonance 
\citep{fabrycky12, veras12, lee13}, with an excess population slightly 
wide of the resonance, and a deficit of planet pairs just inside the 
resonance \citep{lithwick12}.

\citet{marti13} recently showed that the 4:2:1 Laplace resonance in the 
GJ\,876 system \citep{rivera10, baluev11} acts to stabilise the three 
outer planets, constraining their mutual inclinations to less than 20 
degrees and $e_3\ltsimeq$0.05.  \citet{tan13} applied a dynamical 
fitting approach to 10 years of precise Keck radial velocities of the 
HD\,82943 2:1 resonant system \citep{lee06}.  They found a best fit at 
an inclination of 20$\pm$4\degrees\ to the sky plane, which was 
dynamically stable despite the high planetary masses implied by that 
inclination.  Interestingly, \textit{Herschel} debris disk observations 
reported by \citet{kennedy13} show that the disk has a similar 
line-of-sight inclination of 27$\pm$4\degrees.  These examples show how 
planetary systems can be characterized with multiple complementary 
approaches.

HD\,73526 is one of 20 stars added to the Anglo-Australian Planet Search 
(AAPS) in late 1999, based on high metallicity and the then-emerging 
planet-metallicity correlation \citep{laughlin00, vf05}.  The first 
planet, HD\,73526b \citep{tinney03}, was reported to have period 
$P=190.5\pm$3.0 d, eccentricity $e=0.34\pm$0.08, and a minimum mass $m$ 
sin $i=3.0\pm$0.3 \Mjup.  \citet{tinney06} reported a second planet with 
$P=376.9\pm$0.9 days, placing it in the 2:1 resonance with the inner 
planet.  The authors noted that the 2:1 period commensurability appears 
to be common, with 4 of the 18 then-known multiple systems moving on 
such orbits.  The HD 73526 planetary system was reported in a 2:1 mean 
motion resonance (MMR), with $\theta_1$ librating around 0\degrees\ and 
$\theta_2$ circulating \citep{tinney06}, where $\theta_1$ and 
$\theta_2$ are the lowest order, eccentricity-type 2:1 MMR angles:
\begin{equation}
      \theta_1 = \lambda_1 - 2\lambda_2 + \varpi_1  , \\
 \end{equation}
 \begin{equation}
      \theta_2 = \lambda_1 - 2\lambda_2 + \varpi_2  .\\
\end{equation}

\noindent Here, $\lambda$ is the mean longitude, $\varpi$ is the 
longitude of periapse, and subscripts 1 and 2 represent the inner and 
outer planets, respectively.  This type of 2:1 MMR configuration is 
dynamically interesting as it cannot be produced by smooth migration 
capture alone (Beauge et al. 2003; Ferraz-Mello et al. 2003; Lee 2004; 
Beauge et al. 2006; Michtchenko et al. 2008), and alternative mechanisms 
have been suggested to produce such configuration.

The resonant property of the HD\,73526 planets makes this an interesting 
system in terms of its dynamical evolution.  Subsequent work has focused 
on how planets get into the 2:1 resonance in this and other exoplanetary 
systems.  \citet{sandor07} proposed that the HD\,73526 system 
experienced both migration and a sudden perturbation (planet-planet 
scattering or rapid dissipation of the protoplanetary disk) which 
combined to drive the system into the observed 2:1 resonance.  
Similarly, \citet{zhang10} suggested that the HD\,73526 and HD\,128311 
systems, both of which are in 2:1 librating-circulating resonances, 
arrived in that configuration via a hybrid mechanism of scattering and 
collisions with terrestrial planetesimals.  Scattering into low-order 
resonances was also implicated by \citet{raymond08} as a likely 
formation mechanism, where scattering events drive the two larger 
planets into a resonance while ejecting the smaller planet.  In summary, 
there is general agreement that the HD\,73526 system did not arrive in 
the 2:1 resonance by smooth migration alone.


The aim of this work is to provide an updated set of parameters for the 
HD\,73526 system, based on a further 6 years of AAPS observations, as 
well as new data from Magellan (Section 2).  In addition, we perform 
Keplerian and full dynamical fits to the complete data set (Section 3).  
In Section 4, we present detailed dynamical stability maps of the 
system, using both the parameters from the Keplerian and dynamical fits.  
Finally, in Section 5 we offer conclusions on the architecture of the 
system based on the combination of our orbit fitting and dynamical 
stability analysis.

\section{Observations}

\subsection{Anglo-Australian Telescope}

AAPS Doppler measurements are made with the UCLES echelle spectrograph 
\citep{diego:90}.  An iodine absorption cell provides wavelength 
calibration from 5000 to 6200\,\AA.  The spectrograph point-spread 
function and wavelength calibration are derived from the iodine 
absorption lines embedded on every pixel of the spectrum by the cell 
\citep{val:95,BuMaWi96}.  The result is a precision Doppler velocity 
estimate for each epoch, along with an internal uncertainty estimate, 
which includes the effects of photon-counting uncertainties, residual 
errors in the spectrograph PSF model, and variation in the underlying 
spectrum between the iodine-free template, and epoch spectra observed 
through the iodine cell.  All velocities are measured relative to the 
zero-point defined by the template observation.  A total of 36 AAT 
observations have been obtained since 1999 Feb 2 (Table~\ref{AATvels}) 
and used in the following analysis, representing a data span of 4836 
days.  The exposure times range from 300 to 900 sec, and the mean 
internal velocity uncertainty for these data is 4.1\,\ms.

\subsection{Magellan Telescope}

Since HD\,73526 is among the faintest AAPS targets ($V=9.0$), we have 
obtained supplemental observations with the Planet Finder Spectrograph 
(PFS) \citep{crane06,crane08,crane10} on the 6.5m Magellan II (Clay) 
telescope.  The PFS is a high-resolution ($R\sim\,80,000$) echelle 
spectrograph optimised for high-precision radial-velocity measurements 
\citep[e.g.][]{albrecht11, albrecht12, gj667, arr13}.  The PFS also 
uses the iodine cell method, as descibed above, to obtain precise radial 
velocities.  The 20 measurements of HD\,73526 are given in 
Table~\ref{PFSvels}.  The data span 856 days and have a mean internal 
uncertainty of 1.2\,\ms.

\section{Orbit Fitting}

\subsection{Noninteracting Keplerian Fit}

New radial-velocity observations of exoplanetary systems can sometimes 
result in substantial modification of the best-fit planetary orbits.  
For example, the two planets in the HD\,155358 system were initially 
reported to be in orbital periods of 195 and 530 days \citep{cochran07}.  
A further five years of observations revealed that the outer planet 
actually has an orbital period of 391.9 days, and is trapped in the 2:1 
mean-motion resonance \citep{texas1}.  In light of the possibility that 
the best-fit orbits of the two planets may be significantly different 
than initially presented in \citet{tinney06}, we begin our orbit fitting 
process with a wide-ranging search using a genetic algorithm.  This 
approach is often used when the orbital parameters of a planet candidate 
are highly uncertain \citep[e.g.][]{tinney11,HUAqr,NNSer}, or when data 
are sparse \citep{47205paper}.  We allowed the genetic algorithm to 
search a wide parameter space, and it ran for 50,000 iterations, testing 
a total of about $10^7$ possible configurations.  We then fit the two 
data sets simultaneously using \textit{GaussFit} \citep{jefferys87}, a 
generalized least-squares program used here to solve a Keplerian 
radial-velocity orbit model.  The \textit{GaussFit} model has the 
ability to allow the offsets between multiple data sets to be a free 
parameter.  The parameters of the best 2-planet solution obtained by the 
genetic algorithm were used as initial inputs to \textit{GaussFit}, and 
a jitter of 3.3 \ms\ was added in quadrature to the uncertainty of each 
observation (following Tinney et al.~2006).  The best-fit Keplerian 
solutions are given in Table~\ref{planetparams}; planetary minimum 
masses $m$ sin $i$ are derived using a stellar mass of 1.014$\pm$0.046 
\Msun\ \citep{takeda07}.  This fit has a reduced $\chi^2$ of 1.63 and a 
total RMS of 6.32\,\ms\ (AAT -- 7.67\,\ms; PFS -- 2.75\,\ms).

\subsection{Dynamical Fit}

\begin{figure}
\plotone{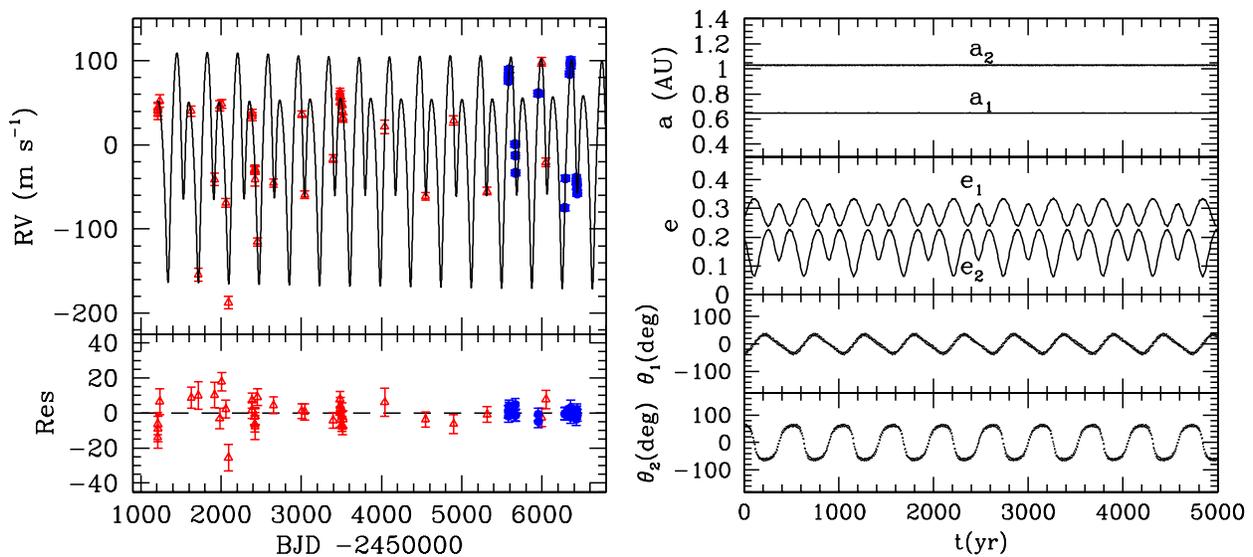}
\caption{Radial-velocity curves and residuals from the coplanar edge-on 
dynamical fit in Table \ref{dynfitresults}.  Error bars include 3.3\ms\ 
of stellar jitter added in quadrature.  Red points are AAT data while 
blue points are PFS data.  The right panel shows the dynamical evolution 
of this system.  The semimajor axes remain essentially constant, while 
the eccentricities show secular variations on timescales of centuries.  
The fit is in a 2:1 MMR with both $\theta_1$ and $\theta_2$ librating 
around 0 degrees.
\label{fig:rv}
}
\end{figure}

Because the two planets are massive enough and orbit close enough to 
each other to be interacting, we also apply a full dynamical model to 
these data.  This model includes the effects of planet-planet 
interactions, and can be used to place constraints on the system's 
inclination to the sky plane, $i$, a quantity which cannot be determined 
from Keplerian fitting alone.  The system inclination then sets the true 
masses of the planets.  The technique is described fully in 
\citet{tan13} for the HD\,82943 two-planet system.  The 
Levenberg-Marquardt (Press et al.~1992) method is adopted as our fitting 
method.  Using the Keplerian best fit as an initial guess, assuming 
coplanar edge-on orbits, the Levenberg-Marquardt algorithm converges to 
a local minimum with $\rchi$ of about 1.70 and RMS of about 6.54 \ms. 
Based on this local minimum, we conduct a parameter grid search 
\citep{lee06, tan13} to ensure a global search for the best fit.  This 
minimum is indeed a global dynamical best fit assuming coplanar edge-on 
orbits; two other local minima with slightly larger $\rchi$ have been 
found.  The coplanar edge-on best-fit parameters are listed in Table 
\ref{dynfitresults}, with their error bars determined by the covariance 
matrix.  This fit and its residuals are shown in the left panel of 
Figure~\ref{fig:rv}.  The right panel of Figure~\ref{fig:rv} 
shows that both resonance angles are librating, with amplitudes of 
$\pm$40\degrees\ ($\theta_1$) and $\pm$60\degrees\ ($\theta_2$).

\begin{figure}
\plotone{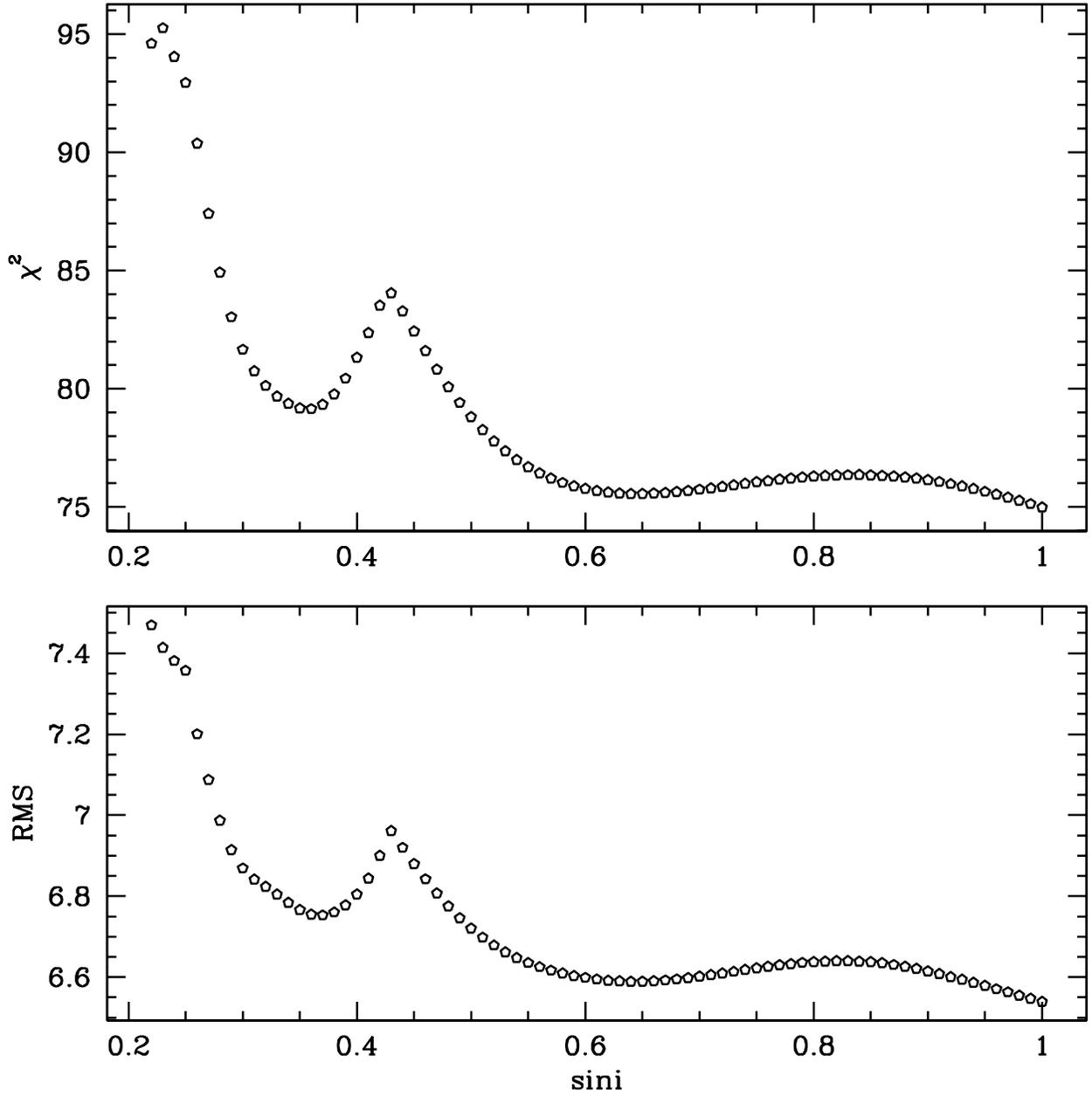}
\caption{  $\chi^2$ and RMS  as a function of $\sin i$.
\label{fig:chisq}
}
\end{figure}
\begin{figure}
\plotone{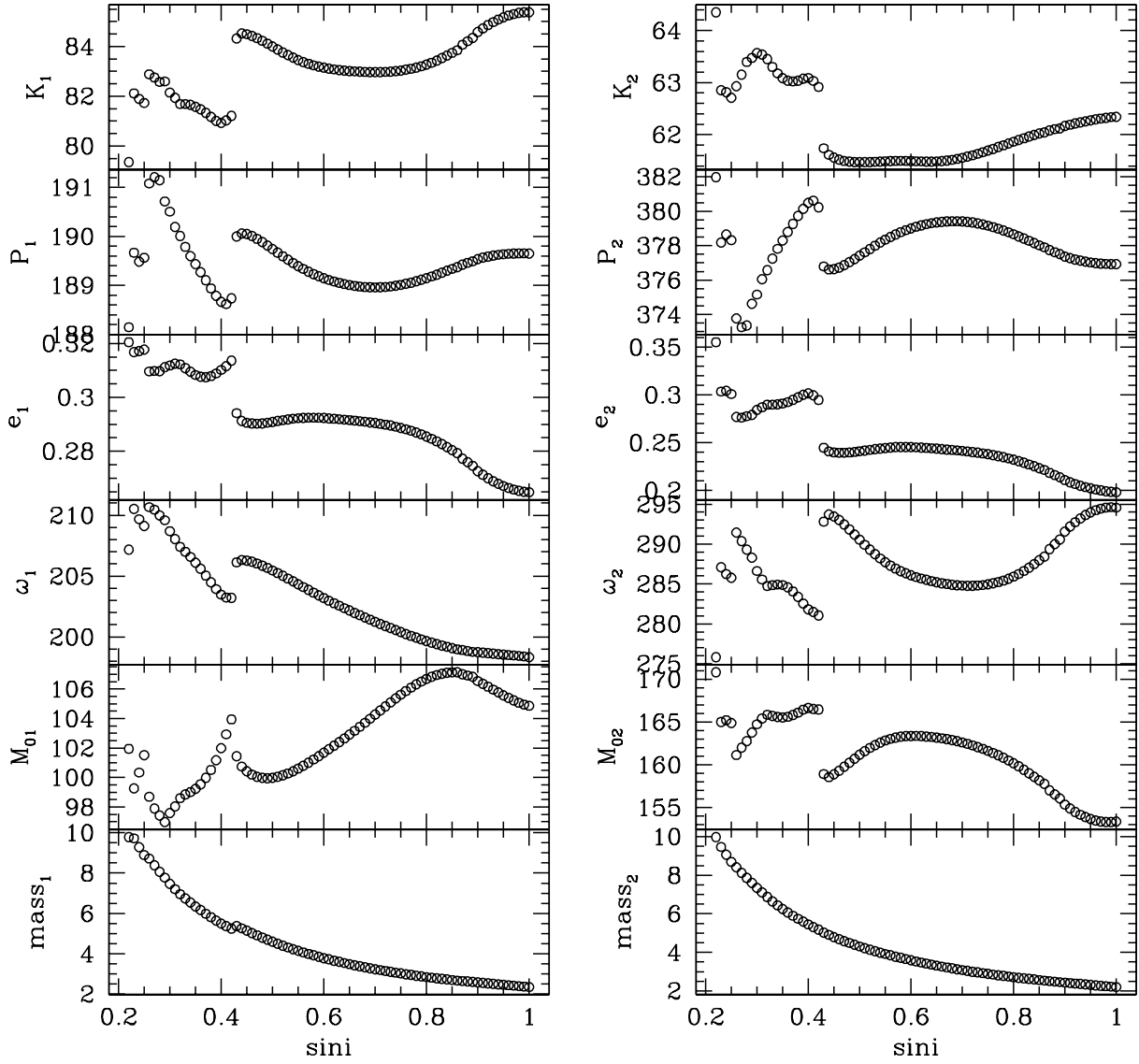}
\caption{  Best-fit parameters as a function of $\sin i$.
\label{fig:para}
}
\end{figure}

Assuming the planets are in coplanar orbits, we then allow the 
inclination to the sky plane to vary along with other fitting 
parameters.  Figure~\ref{fig:chisq} shows $\chi^2$ and RMS as a function 
of $\sin i$, and Figure~\ref{fig:para} shows best-fit parameters as a 
function of $\sin i$.  The $\chi^2$ curve is shallow in the range of 
$\sin i \gtrsim 0.6$, but then shows a clear local minimum at $\sin i 
\sim 0.36$ ($i=20.8$\degrees).  Two further local minima were found, at 
inclinations of $i=$90\degrees\ and 40.2\degrees.  The parameters of 
these three solutions are given in Table~\ref{dynfitresults}; the 
planetary masses scale accordingly as 1/sin\,$i$, resulting in more 
massive planets for the low-inclination solutions.

\begin{figure}
\plotone{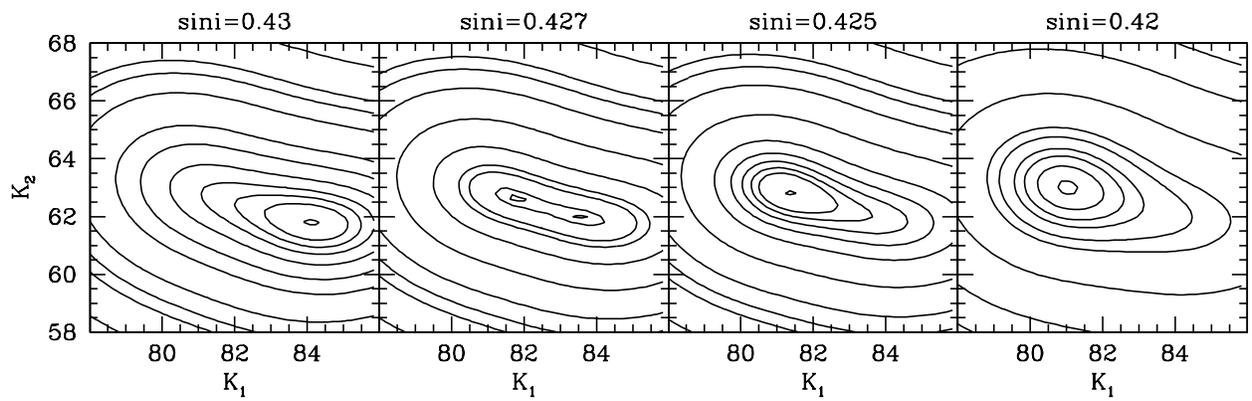}
\caption{ Evolution of $\chi^2_{\nu}$ contours in $K_1$-$K_2$ space with 
as a function of $\sin i$.
\label{fig:K1K2}
}
\end{figure}

The $\chi^2$ curve and fitting parameters (Fig.~\ref{fig:para}) show 
discontinuities along the $\sin i$ axis, especially those near $\sin i 
\sim 0.43$.  To understand these discontinuities, we explore grids in 
different fixed $\sin i$, to see the evolution of the parameter space 
along different inclinations.  Figure~\ref{fig:K1K2} shows $K_1$-$K_2$ 
grids for different $\sin i$.  Initially when the orbits are at $\sin i 
\sim 0.43$, there is only one $\chi^2$ minimum ($K_1\sim 84$).  Then a 
new local minimum ($K_1 \sim 82 $) appears around $\sin i = 0.427$, 
whose fitting parameters are significantly different from the original 
minimum ($K_1 \sim 84$).  When $\sin i$ drops down to 0.425, the 
original minimum vanishes and the new one becomes a single minimum in 
parameter space.  The appearance of the additional $\rchi$ minimum 
results in the big ``jump'' of fitting parameters at about $\sin i = 
0.43$ (see Fig.~\ref{fig:para}).


In summary, we have four possible configurations for this system (one 
Keplerian fit and three dynamical fits).  The four solutions are not 
substantially different from one another, apart from the sin~$i$ factor 
for the three solutions in Table~\ref{dynfitresults}, which serves to 
increase the planetary masses relative to the Keplerian scenario in 
which we have assumed the planets to be at their minimum masses 
(m~sin~$i$).  As a first-order check of dynamical stability, 
the best-fit system configuration at each inclination was integrated for 
$10^8$ yr.  For all fits so tested, at inclinations of 26.7, 30.0, 33.4, 
36.9, 40.0, and 90.0 degrees (sin\,$i=$0.45, 0.5, 0.55, 0.6, 0.64, 1.0), 
the systems remained stable for $10^8$ yr.  However, since dynamical 
stability is highly dependent on the initial conditions, we expand on 
these tests in the next section to obtain a more robust and complete 
picture of the stability of the various configurations.

\section{Dynamical Stability Testing}

We have found four possible solutions for the HD\,73526 system, which 
significantly differ in inclination (and hence the planetary masses).  
It is therefore critical to perform dynamical stability tests on these 
configurations, as the solution with the absolute $\chi^2$ minimum may 
prove dynamically unfeasible.

\subsection{Long-Term Stability}

When analyzing any multiple-planet system, it is prudent to investigate 
the long-term dynamical stability of the system.  As more multi-planet 
systems are discovered, the announcement of planetary systems which turn 
out to be dynamically unfeasible has become increasingly common.  
Detailed N-body simulations can be used to test the veracity of planet 
claims.  Sometimes the results of such tests have shown that some 
systems simply cannot exist in their proposed configuration on 
astronomically relevant timescales \citep[e.g.][]{horner11, HUAqr, 
HWVir, NSVS}.  In other cases, dynamical testing can place additional 
constraints on planetary systems, particlarly when the planets are in or 
near resonances \citep[e.g.][]{texas1, texas2, subgiants}.  In this 
section, we examine the various solutions for the HD\,73526 system, 
performing detailed dynamical tests of the planetary system 
configurations as given in Tables~\ref{planetparams} and 
\ref{dynfitresults}.  Given that the four solutions are not 
substantially different from each other in terms of goodness-of-fit, 
these dynamical stability tests can serve to discern which scenario is 
most plausible: a solution which is favored by the fitting process may 
prove to be unstable and hence unfeasible.

\subsection{Procedure}

As in our previous dynamical work \citep[e.g. 
][]{marshall10,142paper,NNSer}, we used the Hybrid integrator within the 
\textit{N}-body dynamics package {\sc Mercury} \citep{chambers99} to 
perform our integrations.  We held the initial orbit of the inner planet 
fixed at its best-fit parameters, as given in Table~\ref{planetparams}, 
and then created 126,075 test systems.  In those test systems, the 
initial orbit of the outer planet was varied systematically in 
semi-major axis $a$, eccentricity $e$, periastron argument $\omega$, and 
mean anomaly $M$, resulting in a 41x41x15x5 grid of ``clones'' spaced 
evenly across the 3$\sigma$ range in those parameters.  We assumed the 
planets were coplanar with each other and, for the Keplerian case, we 
assigned masses equivalent to their minimum mass, $m\sin i$ 
(Table~\ref{planetparams}).  We then followed the dynamical evolution of 
each test system for a period of 100 million years, and recorded the 
times at which either of the planets was removed from the system.  
Planets were removed if they collided with one another, hit the central 
body, or reached a barycentric distance of 10~AU.

We performed these dynamical simulations for the Keplerian fit 
($i=90$\degrees), the dynamical fit at $i=90$\degrees, and the 
lowest-inclination dynamical fit: the configuration given in 
Table~\ref{dynfitresults} at $i=20.8$\degrees.  For the latter scenario, 
the planet masses were scaled according to the derived system 
inclination $i$.  Clearly, the masses of the planets are a proxy for the 
expected dynamical stability -- systems containing more-massive planets 
are likely to be less stable.  Hence, the three scenarios we have 
tested, at $i=90$\degrees\ and $i=20.8$\degrees, represent the extremes 
of dynamical stability (or instability) for the HD\,73526 system.

To explore the effects of mutual inclinations between the planets, we 
performed five additional N-body simulations, for scenarios in which the 
two planets were inclined with respect to each other.  These simulations 
were set up exactly as described above, using the parameters of the 
Keplerian solution (Table~\ref{planetparams}), except at a lower 
resolution due to computing limitations: a 21x21x5x5 grid in $a$, $e$, 
$\omega$, and $M$.  Five runs were performed, at mutual inclinations of 
5, 15, 45, 135, and 180 degrees.  The latter two cases represent 
retrograde scenarios, which can sometimes allow for a larger range of 
dynamically stable orbits \citep{eberle10, horner11}.

\subsection{Results}

The results of our dynamical stability simulations for the Keplerian 
solution are shown in Figure~\ref{dynam1}.  We show six panels, for the 
coplanar and five mutually-inclined scenarios as described above.  For 
the coplanar and 5-degree cases (panels a and b), the best-fit set of 
parameters (shown by the open square with 1$\sigma$ crosshairs) lies in 
a region of moderate stability, with mean system survival times of 
$\sim\,10^{6}$ years.  The stability rapidly degrades as the inclination 
between the planets becomes significant, and even for retrograde cases 
(panels e and f), the nominal best-fit system destabilizes within $10^4$ 
yr.  From these simulations, we can conclude that the HD\,73526 planets 
are most likely coplanar with each other.  Panels (a) and (b) also show 
that the stability of the system increases as the outer planet takes on 
lower eccentricities.  For $e\ltsimeq$0.2, mean survival times exceed 
$10^7$ yr.  This is not a surprising result, as high eccentricities 
generally increase the possibility of strong interactions or 
even orbit crossings (though systems in protected resonances 
may remain stable for some values of $M$ and $\omega$).  Indeed, the 
statistics of multi-planet systems show that planets in multiple systems 
tend to have lower eccentricities \citep{wright09, wittenmyer09}.

As shown in Figure~\ref{dynam1}, the Keplerian best-fit solution 
is stable on million-year timescales.  However, the colored squares in 
Figure~\ref{dynam1} represent the \textit{mean} survival times across 
the range of mean anomalies and $\omega$ tested.  As the best-fit 
solution for the HD\,73526 system places the planets on resonant orbits, 
their stability will naturally be highly sensitive to the values of 
these angles.  Hence, Figure~\ref{angles} shows the outcomes of the 75 
individual simulations performed at the best-fit $a$ and $e$ spanning a 
5x15 grid in $M$ and $\omega$.  We see that the nominal solution (where 
$a$ and $e$ are fixed at the best-fit values) lies at the point of 
maximum stability.  While this is reassuringly consistent with our 
expectations of enhanced stability within the resonance, we caution that 
each colored square in Figure~\ref{angles} represents only a single run, 
and dynamical evolution is known to be a chaotic process 
\citep[e.g.][]{horner04a,horner04b}.

The long-term stability results for the dynamical fit with 
$i=90$\degrees\ are shown in Figure~\ref{dynam3}.  It is immediately 
apparent that this solution results in a higher degree of stability, 
with the entire one-sigma region exhibiting mean lifetimes exceeding 
$10^7$\,yr.  The right panel of Figure~\ref{dynam3} shows the results 
from the 75 individual runs in the central best-fit square, as in 
Figure~\ref{angles}.  For this case, when we examine the dependence on 
$M$ and $\omega$, we see that the entire diagonal region (including the 
best fit) is stable for $10^8$ years.  In contrast, Figure~\ref{dynam2} 
shows the dynamical stability of the $i=20.8$\degrees\ solution from 
Table~\ref{dynfitresults}.  Though this fit is formally almost as good 
as the Keplerian fit, the increased masses implied by the inclination 
render the system unstable on short timescales (1000 yr).


\section{Discussion and Conclusions}

We have fit the HD\,73526 system using both kinematic and dynamical 
techniques, yielding four possible solutions.  There are no compelling 
differences between the four models in terms of their goodness-of-fit 
statistics or derived planetary parameters.  The only significant 
distinguishing characteristics are the planet masses derived from the 
system inclinations in the dynamical fits (Table~\ref{dynfitresults}).  
We thus turned to a detailed dynamical stability mapping procedure in 
which we tested a broad range of parameters about the best-fit 
solutions.

\begin{figure}
\plotone{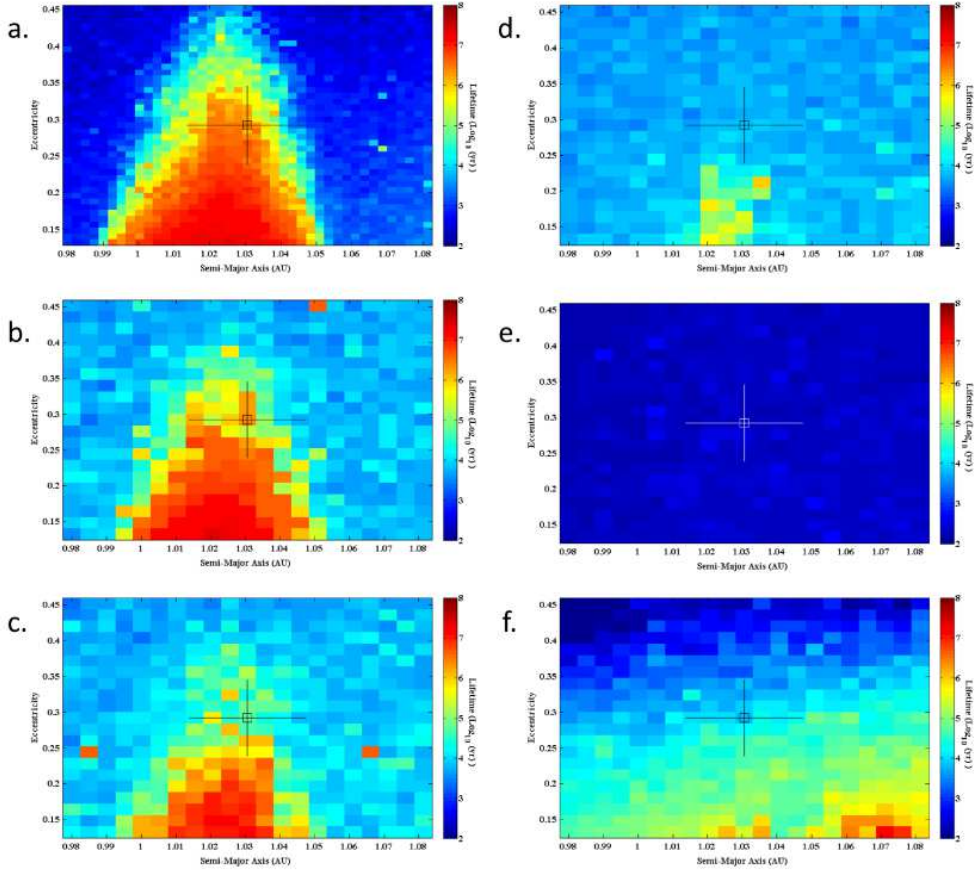}
\caption{Dynamical stability of the HD\,73526 system as a function of 
the outer planet's initial eccentricity and semimajor axis.  The 
best-fit Keplerian parameters (Table~\ref{planetparams}) are marked by 
the open box with 1$\sigma$ crosshairs.  Each colored square represents 
the mean lifetime of 75 unique $M$-$\omega$ combinations at that point 
in ($e$,$a$) for the outer planet.  Panel (a) is the coplanar case, and 
panels (b)-(f) are the mutually-inclined scenarios, for inclinations of 
5,15,45,135, and 180 degrees, respectively. }
\label{dynam1}
\end{figure}

\begin{figure}
\plotone{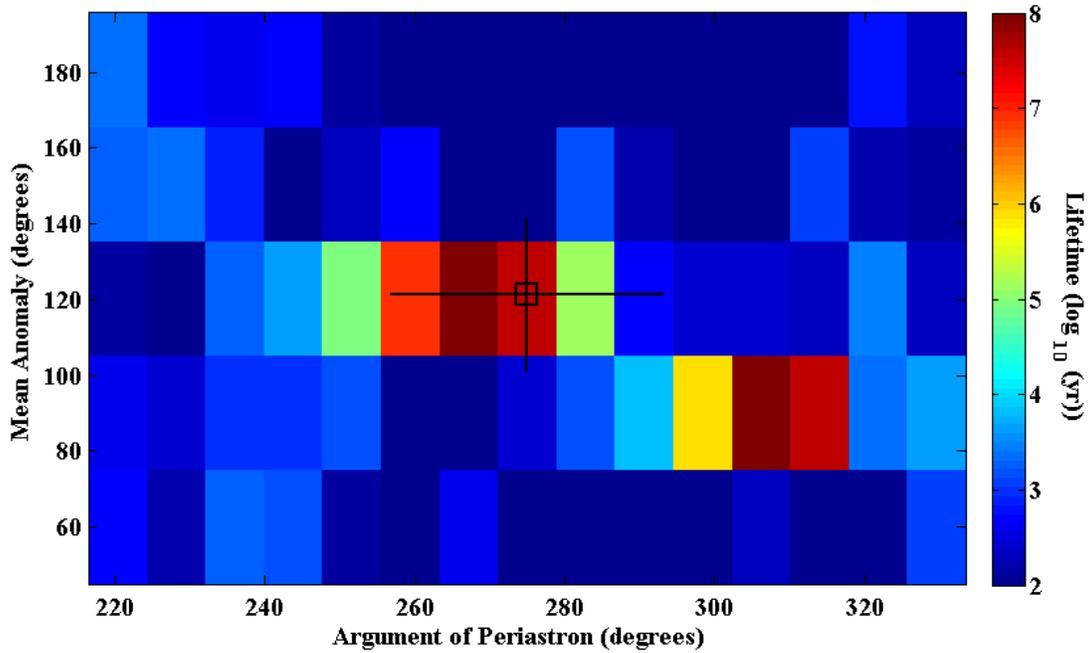}
\caption{Dynamical stability of the best-fit Keplerian solution for the 
HD\,73526 system for a 15x5 grid of $\omega$ and $M$.  The semimajor 
axis and eccentricity have been fixed to their best-fit values.  The 
colors and symbols have the same meaning as in Figure~\ref{dynam1}; this 
plot shows results from the 75 individual simulations which were 
averaged in the center colored square of Figure~\ref{dynam1}.  The 
best-fit solution lies squarely in the most stable region of this subset 
of simulations.}
\label{angles}
\end{figure}

\begin{figure}
\plottwo{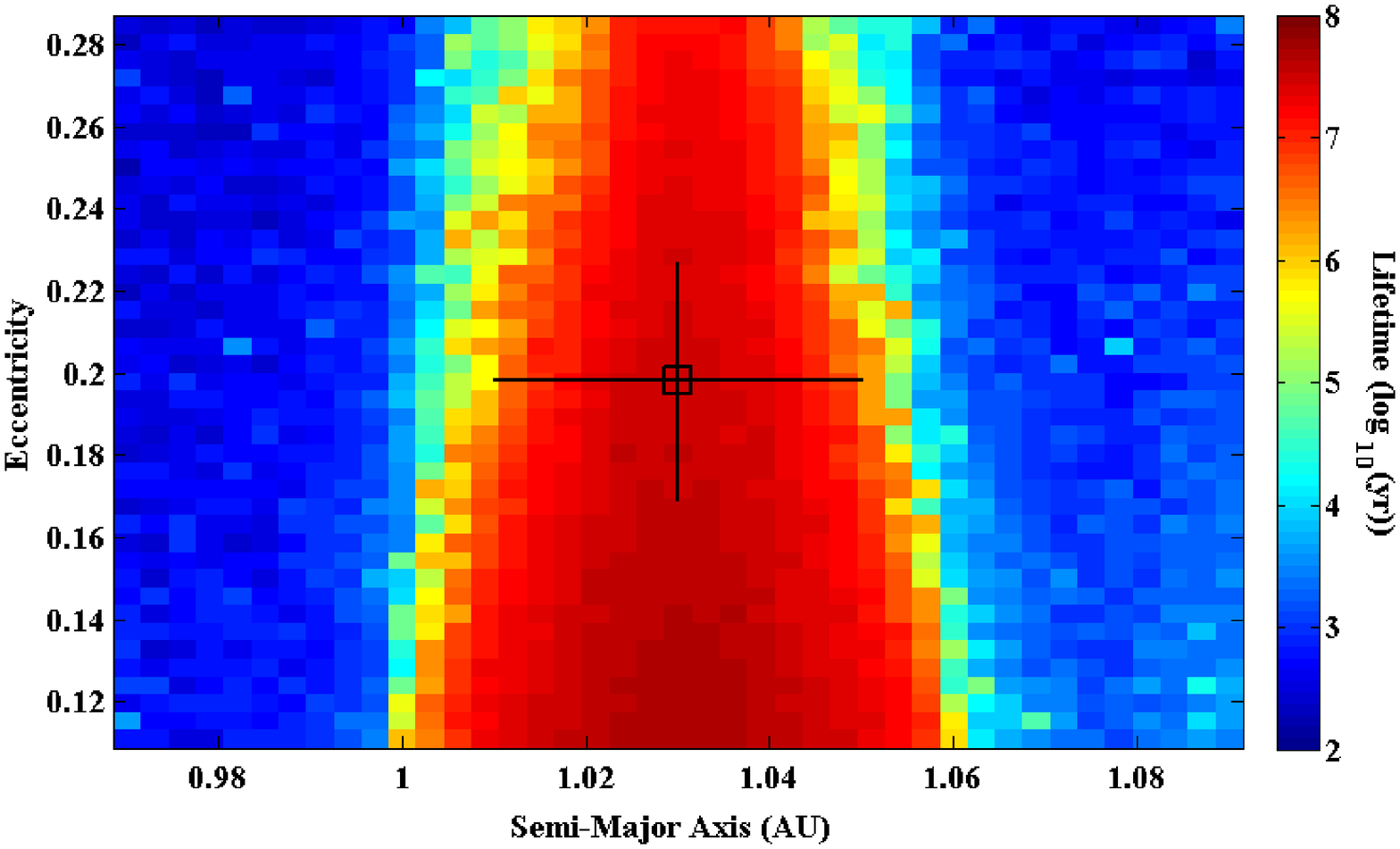}{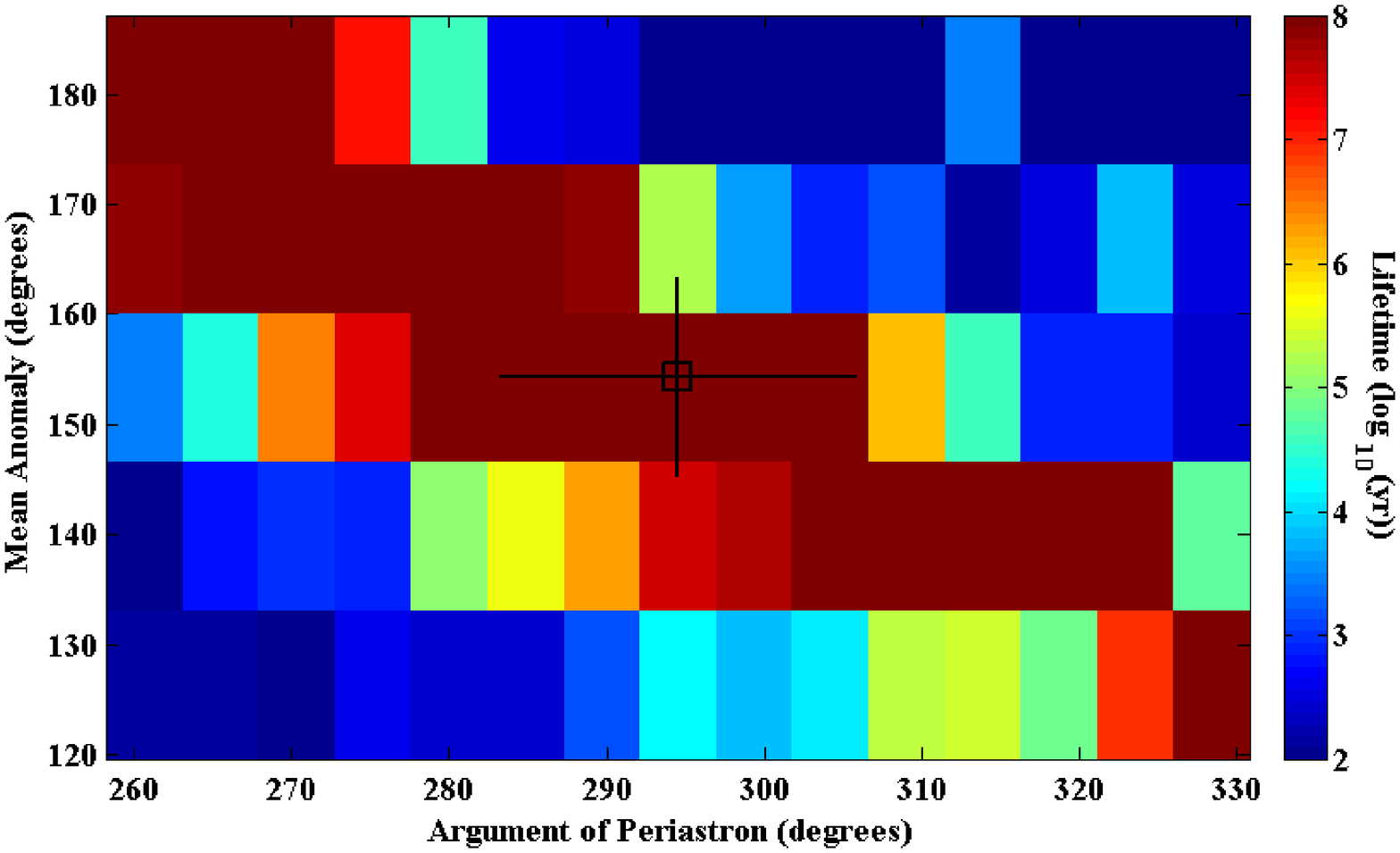}
\caption{Left: Stability of the HD\,73526 system as a function of the 
outer planet's initial eccentricity and semimajor axis.  The colors and 
symbols have the same meaning as in Figure~\ref{dynam1}.  For this 
system, we used the $i=90$\degrees\ solution 
(Table~\ref{dynfitresults}).  As compared to the Keplerian solution, 
this fit results in substantially enhanced stability throughout the 
$1\sigma$ range. Right: Same as Figure~\ref{angles}, but for the 
dynamical-fit $i=90$\degrees\ solution. }
\label{dynam3}
\end{figure}

\begin{figure}
\plotone{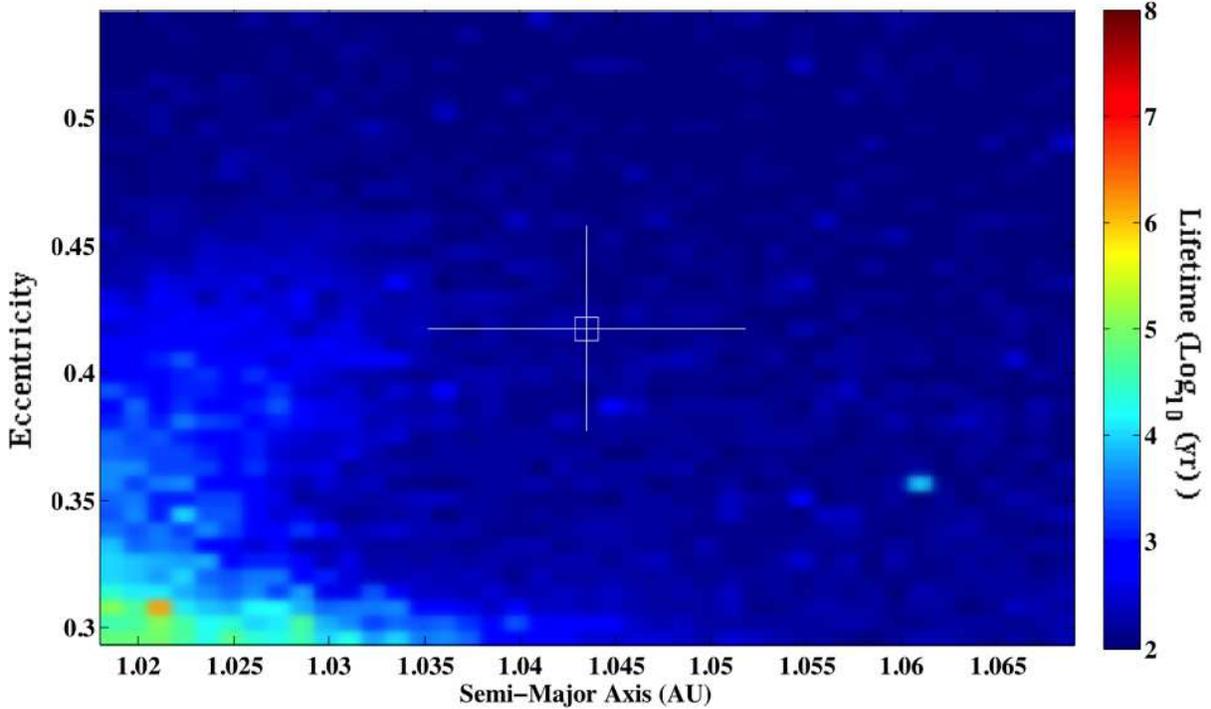}
\caption{Same as Figure~\ref{dynam3}, but for the $i=20.8$\degrees\ 
solution (Table~\ref{dynfitresults}), resulting in higher masses for the 
planets.  Hence, the system is much less stable than the minimum-mass 
case explored in Figure~\ref{dynam3}, with mean survival times of order 
1000 years. }
\label{dynam2}
\end{figure}

Our dynamical stability testing showed that the Keplerian model yielded 
a system which was stable on million-year time scales, with stability 
increasing for lower eccentricities (Figure~\ref{dynam1}).  The 
interacting dynamical fitting procedure gave three ``best'' solutions, 
one of which was at a system inclination of 90\degrees\ (giving planet 
masses equal to the m sin $i$ minimum masses used in the Keplerian 
model).  Our stability testing for the inclined solution at 
$i=20.8$\degrees\ resulted in severe instability throughout the allowed 
3$\sigma$ parameter space.  The increased planetary masses for the 
low-inclination solutions appear to destabilize the system on 
astronomically short timescales ($<$1000 yr).  This result leads us to 
reject the $i=20.8$\degrees\ scenario.  While the individual best-fit 
solutions proved stable for $i>26.7$\degrees (as noted in Section 3.2), 
it is clear that the region of long-term stability expands as the system 
inclination increases.  We thus adopt the $i=90$\degrees\ dynamical fit 
for two primary reasons: first, the planets are massive enough that they 
are certainly interacting with each other, as evidenced by the 2:1 
resonant configuration; and second, this fit proved to be significantly 
more stable than the Keplerian fit (Figure~\ref{dynam1}).  We note in 
passing that if the system's inclination is indeed near 90\degrees, 
there is the possibility that one or both planets transit.

This work has shown how dynamical stability considerations can serve to 
constrain the configuration of a planetary system when the $\chi^2$ 
surface is such that a clear minimum is not evident 
\citep[e.g.][]{campanella11}.  We have combined two fitting methods with 
the detailed dynamical simulations to present an updated view of the 
interesting 2:1 resonant planetary system orbiting HD\,73526.

\acknowledgements

This research has made use of NASA's Astrophysics Data System (ADS), and 
the SIMBAD database, operated at CDS, Strasbourg, France.  This research 
has also made use of the Exoplanet Orbit Database and the Exoplanet Data 
Explorer at exoplanets.org \citep{wright11}.  M.H.L. was supported in 
part by the Hong Kong RGC grant HKU 7024/13P.  DM acknowledges funding 
from the BASAL CATA Center for Astrophysics and Associated Technologies 
PFB-06, and the The Milky Way Millennium Nucleus from the Ministry for 
the Economy, Development, and Tourism's Programa Iniciativa Cient\'ifica 
Milenio P07-021-F.



\begin{deluxetable}{lrr}
\tabletypesize{\scriptsize}
\tablecolumns{3}
\tablewidth{0pt}
\tablecaption{AAT/UCLES Radial Velocities for HD 73526}
\tablehead{
\colhead{JD-2400000} & \colhead{Velocity (\ms)} & \colhead{Uncertainty
(\ms)}}
\startdata
\label{AATvels}
51212.13020  &      7.91  &    5.40  \\
51213.13145  &      0.32  &    5.37  \\
51214.23895  &      5.52  &    6.54  \\
51236.14647  &     15.70  &    6.62  \\
51630.02802  &      3.91  &    5.12  \\
51717.89996  &   -190.60  &    7.09  \\
51920.14186  &    -77.28  &    6.45  \\
51984.03780  &     10.04  &    4.87  \\
52009.09759  &     12.38  &    4.21  \\
52060.88441  &   -105.26  &    3.73  \\
52091.84653  &   -223.76  &    6.76  \\
52386.90032  &     -2.62  &    3.46  \\
52387.89210  &      1.72  &    2.86  \\
52420.92482  &    -66.78  &    3.28  \\
52421.91992  &    -64.87  &    3.23  \\
52422.86019  &    -66.65  &    3.34  \\
52424.92369  &    -77.31  &    7.02  \\
52454.85242  &   -151.57  &    3.44  \\
52655.15194  &    -81.59  &    3.53  \\
53008.13378  &      0.13  &    2.42  \\
53045.13567  &    -95.56  &    3.20  \\
53399.16253  &    -52.76  &    2.97  \\
53482.87954  &     20.95  &    2.02  \\
53483.88740  &     26.55  &    2.59  \\
53485.96240  &     22.83  &    3.65  \\
53488.93814  &     14.81  &    2.33  \\
53506.88650  &      5.03  &    2.25  \\
53508.91266  &     11.94  &    2.03  \\
53515.89441  &     -4.01  &    2.89  \\
53520.91025  &     -4.97  &    3.27  \\
54041.18613  &    -14.76  &    7.28  \\
54549.03413  &    -97.63  &    2.83  \\
54899.03133  &     -7.35  &    4.24  \\
55315.92532  &    -91.43  &    3.06  \\
55997.03979  &     62.28  &    4.10  \\
56048.94441  &    -57.19  &    4.16  \\
\enddata
\end{deluxetable}


\begin{deluxetable}{lrr}
\tabletypesize{\scriptsize}
\tablecolumns{3}
\tablewidth{0pt}
\tablecaption{Magellan/PFS Radial Velocities for HD 73526}
\tablehead{
\colhead{JD-2400000} & \colhead{Velocity (\ms)} & \colhead{Uncertainty
(\ms)}}
\startdata
\label{PFSvels}
55582.79672  &     14.7  &    1.2  \\
55584.75698  &     20.6  &    1.2  \\
55585.74045  &     22.1  &    1.2  \\
55587.77487  &     28.3  &    1.0  \\
55588.71850  &     28.4  &    0.9  \\
55663.53102  &    -60.3  &    1.1  \\
55668.54537  &    -73.8  &    0.8  \\
55672.50855  &    -94.5  &    0.8  \\
55953.76750  &      0.4  &    1.3  \\
55955.71181  &      0.0  &    1.1  \\
56282.77476  &   -135.9  &    1.5  \\
56292.76731  &   -101.0  &    1.3  \\
56345.67804  &     23.1  &    1.2  \\
56355.63611  &     33.8  &    1.3  \\
56357.65331  &     33.0  &    1.2  \\
56358.70107  &     39.4  &    2.4  \\
56428.46819  &    -99.9  &    1.2  \\
56431.48616  &   -105.3  &    1.7  \\
56434.49819  &   -110.7  &    1.1  \\
56438.46472  &   -119.0  &    1.1  \\
\enddata
\end{deluxetable}

\begin{deluxetable}{lr@{$\pm$}lr@{$\pm$}lr@{$\pm$}lr@{$\pm$}lr@{$\pm$}lr@{$\pm$}
lr@{$\pm$}l}
\tabletypesize{\scriptsize}
\tablecolumns{10}
\tablewidth{0pt}
\tablecaption{Keplerian Orbital Solutions }
\tablehead{
\colhead{Planet} & \multicolumn{2}{c}{Period} & \multicolumn{2}{c}{$T_0$}
&
\multicolumn{2}{c}{$e$} & \multicolumn{2}{c}{$\omega$} &
\multicolumn{2}{c}{K } & \multicolumn{2}{c}{m sin $i$ } &
\multicolumn{2}{c}{$a$ } \\
\colhead{} & \multicolumn{2}{c}{(days)} & \multicolumn{2}{c}{(JD-2400000)}
&
\multicolumn{2}{c}{} &
\multicolumn{2}{c}{(degrees)} & \multicolumn{2}{c}{(\ms)} &
\multicolumn{2}{c}{(\Mjup)} & \multicolumn{2}{c}{(AU)}
 }
\startdata
\label{planetparams}   
HD 73526 b & 188.9 & 0.1 & 52856 & 2 & 0.29 & 0.03 & 196 & 5 &
82.7 & 2.5 & 2.25 & 0.12 & 0.65 & 0.01 \\
HD 73526 c & 379.1 & 0.5 & 53300 & 10 & 0.28 & 0.05 & 272 & 10 &
65.1 & 2.6 & 2.25 & 0.13 & 1.03 & 0.02 \\
\enddata
\end{deluxetable}

\begin{deluxetable}{lrr}
\tabletypesize{\scriptsize}
\tablecolumns{3}
\tablewidth{0pt}
\tablecaption{Dynamical Fit Solutions}
\tablehead{
\colhead{Parameter} & \colhead{Planet b} & \colhead{Planet c}}
\startdata
\label{dynfitresults}
$K$ [\ms] & 85.4$\pm$2.3 & 62.3$\pm$1.8 \\
Period [days] & 189.65$\pm$0.21 & 376.93$\pm$0.69 \\ 
Eccentricity & 0.265$\pm$0.021 & 0.198$\pm$0.029 \\
$\omega$ [deg] & 198.3$\pm$3.6 & 294.5$\pm$11.3 \\
Mean anomaly [deg] & 105.0$\pm$5.0 & 153.4$\pm$9.0 \\
$a$ [AU] &  0.65$\pm$0.01 & 1.03$\pm$0.02 \\
$i$ [deg] & 90.0 & \\
Mass [\Mjup] & 2.35$\pm$0.12 & 2.19$\pm$0.12 \\
$\chi^{2}_{\nu}$ & 1.70 & \\
RMS [\ms] & 6.54 & \\
\hline
$K$ [\ms] & 83.0$\pm$2.1 & 61.5$\pm$1.6 \\
Period [days] & 189.01$\pm$0.27 & 379.32$\pm$0.92 \\
Eccentricity & 0.292$\pm$0.022 & 0.244$\pm$0.026 \\
$\omega$ [deg] & 202.3$\pm$3.2 & 285.3$\pm$10.6 \\
Mean anomaly [deg] & 102.8$\pm$3.9 & 163.2$\pm$8.3 \\
$a$ [AU] & 0.65$\pm$0.01 & 1.03$\pm$0.02 \\
$i$ [deg] & 40.2 & \\
Mass [\Mjup] & 3.50$\pm$0.17 & 3.32$\pm$0.17 \\
$\chi^{2}_{\nu}$ & 1.72 & \\
RMS [\ms] & 6.59 & \\
\hline
$K$ [\ms] & 81.4$\pm$2.3 & 63.1$\pm$1.6 \\
Period [days] & 189.43$\pm$0.82 & 378.29$\pm$2.81 \\
Eccentricity & 0.308$\pm$0.020 & 0.293$\pm$0.021 \\
$\omega$ [deg] & 205.7$\pm$3.4 & 284.3$\pm$9.9 \\
Mean anomaly [deg] & 99.5$\pm$4.0 & 165.8$\pm$6.8 \\
$a$ [AU] & 0.649$\pm$0.012 & 1.029$\pm$0.021 \\
$i$ [deg] & 20.8 & \\
Mass [\Mjup] & 6.22$\pm$0.33 & 6.10$\pm$0.31 \\
$\chi^{2}_{\nu}$ & 1.80 & \\
RMS [\ms] & 6.76 & \\
\enddata
\end{deluxetable}


\end{document}